\documentclass{eas}
\usepackage{graphicx}
\usepackage{natbib}

\graphicspath{{pic/}}

\begin{document}

\title{The convection of close red supergiant stars observed with near-infrared interferometry}
\runningtitle{Interferometric observations of close red supergiants}
\author{M.~Montarg\`es}\address{Institut de Radioastronomie Millim\'etrique, 300 rue de la Piscine, 38406, Saint Martin d'H\`eres, France}\secondaddress{LESIA, Observatoire de Paris, PSL Research University, CNRS UMR 8109, UPMC, Universit\'e Paris Diderot}
\author{P.~Kervella}\address{Unidad Mixta Internacional Franco-Chilena de Astronom\'{i}a (UMI 3386), CNRS/INSU, France \& Departamento de Astronom\'{i}a, Universidad de Chile}\sameaddress{, 2}
\author{G.~Perrin}\sameaddress{2}
\author{A.~Chiavassa}\address{Laboratoire Lagrange, Universit\'e C\^ote d'Azur, Observatoire de la C\^ote d'Azur, CNRS}
\author{M.~Auri\`ere}\address{Universit\'e de Toulouse, UPS-OMP, Institut de Recherche en Astrophysique et Plan\'etologie, Toulouse, France}
%
%
\begin{abstract}
Our team has obtained observations of the photosphere of the two closest red supergiant stars Betelgeuse ($\alpha$  Ori)  and Antares  ($\alpha$  Sco) using near infrared interferometry.
We have been monitoring the photosphere of Betelgeuse with the VLTI/PIONIER instrument for three years. On Antares, we obtained an unprecedented sampling of the visibility function. These data allow us to probe the convective photosphere of massive evolved stars.
\end{abstract}
\maketitle

\section{Introduction}

Chemical evolution of the Universe is driven by evolved stars. During their red supergiant (RSG) stage, massive stars have an important mass loss. The process at the origin of this outflow remains unknown. Indeed, RSG do not experience large pulsations or flare that could input the requested momentum.
From spectroscopic observations of some RSG, \citet{2007A&A...469..671J} suggested that lowering of the effective gravity caused by the convective cells predicted by \citet{1975ApJ...195..137S} together with radiative pressure on molecular lines could trigger the outflow. These structures were later observed by \citet{2009A&A...508..923H}, \citet{2010A&A...515A..12C}, and \citet{2014A&A...572A..17M} on the surface of Betelgeuse. 

Interferometric observations of nearby RSG are the best opportunity to probe the photosphere of these stars.

\section{Observations}

We observed the nearby RSG Antares ($\alpha$~Sco, HD~148478, HR~6134) and Betelgeuse ($\alpha$~Ori, HD~39801, HR~2061). With their large angular diameters, they are the best suited targets for a detailed study of their close circumstellar environment (CSE). We observed these stars using the PIONIER instrument \citep{2011A&A...535A..67L} which recombines the light of 4 telescopes of the VLTI \citep{2010SPIE.7734E...3H} in the H band at low spectral resolution. 

Antares was observed with the three available configurations of the auxiliary telescopes (AT) of 1.8~m of diameter: A1-B2-C1-D0 (2014 April 22), D0-H0-G1-I1 (2014 April 28) and A1-G1-K0-J3 (2014 May 03 and 06). The baseline lengths on the ground were ranging from to 11.3~m to 153.0~m.

The observations of Betelgeuse were performed only with the compact quadruplet A1-B2-C1-D0 on the nights of  2012 January 31, 2013 February 09, 2014 January 11, 2014 February 01 and 2014 November 21. The baseline lengths were ranging from 11.3 to 35.8~m.

The data were reduced using the PIONIER pipeline (\texttt{pndrs}, \citealt{2011A&A...535A..67L}).

\section{Data analysis}

\subsection{Up to a resolution of 3~mas on Antares}

\begin{figure}[!ht]
	\centering
	\resizebox{\hsize}{!}{\includegraphics{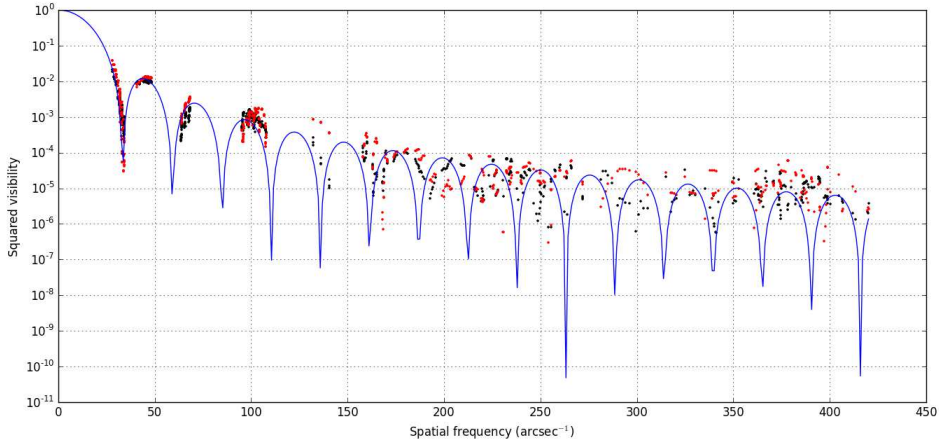}}
	\caption{Squared visibilities on Antar\`es. Black: PIONIER observations. Blue: Best fit LDD model on spatial frequencies lower than 50 arcsec$^{-1}$. Red: Best match RHD simulation.}
	\label{Fig:Antares_Visi}
\end{figure}

The data obtained on Antares allow us to obtain the unprecedented resolution of less than one tenth of the stellar diameter. The angular size of the star was determined with a fit of a power law limb-darkened disk (LDD) model \citep{1997A&A...327..199H} on the first two lobes of the visibility function (spatial frequencies lower than 50 arcsec$^{-1}$). We derived a LDD diameter $\theta_\mathrm{LDD} = 39.79 \pm 1.11$~mas and a power law coefficient $\alpha_\mathrm{LDD} = 0.66 \pm 0.19$ for $\chi^2 = 16$ (reduced). The result of this fit is presented on Fig. \ref{Fig:Antares_Visi}. Our values are slightly different from previous measurements of \citet{2013A&A...555A..24O} but their observations were in the K band. At high spatial frequencies, the observed visibilities are higher than what the LDD model predicts: this indicates the presence of small scale structures.

To explore the higher spatial frequencies domain, following the method of \citet{2010A&A...515A..12C}, and \citet{2014A&A...572A..17M}, we used radiative-hydrodynamics (RHD) simulations obtained with the CO$^5$BOLD code \citep{2012JCoPh.231..919F}. Several temporal snapshots of the simulations were used to give a sample of convection pattern. Intensity maps were computed using the Optim3d code \citep{2009A&A...506.1351C}. Each intensity map was also rotated around its center, at 18 positions between 0$^\circ$ and 180$^\circ$ to compensate the indeterminacy of the orientation relatively to the star. We finally took the Fourier transform of each intensity image to obtain the interferometric observables (more of details on this process are available in Chiavassa et al. in this proceeding). 

The simulation st36g00n05 \citep{2011A&A...535A..22C} matches our visibilities: although the values are not exactly reproduced (reduced $\chi^2 = 158$ , and see Fig. \ref{Fig:Antares_Visi}), the simulation is in the same range of values above the LDD model. This indicates that we have matching characteristic sizes for the convective cells but not in the same pattern as it was on the star at the time of the observations.

\subsection{Monitoring of the photosphere of Betelgeuse}

\begin{figure}[!ht]
	\centering
	\resizebox{\hsize}{!}{\includegraphics{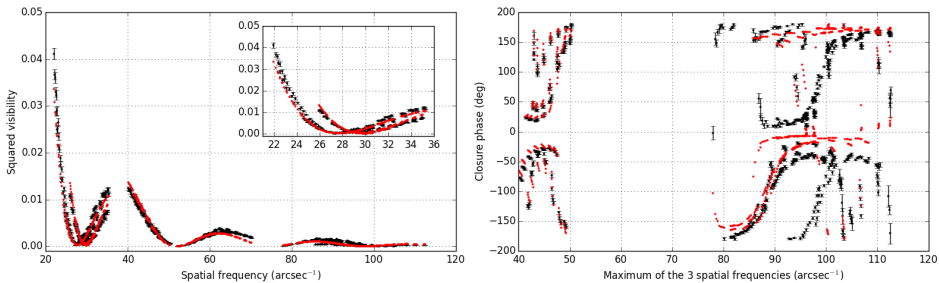}}
	\caption{Fit of the PIONIER squared visibilities (\textit{left}) and closure phases (\textit{right}) observed in February 2013 by the LDD and gaussian hot spot model. Only spatial frequencies lower than 51 arcsec$^{-1}$ were considered. The data are in black and the best fit model in red.}
	\label{Fig:v2_cp_hotspot_2013}
\end{figure}

The observed squared visibilities  present an azimuthal dependence of the spatial frequency at which the first zero occurs, thus a position angle (PA) dependence of the stellar radii (Fig. \ref{Fig:v2_cp_hotspot_2013}). This means that neither a uniform disk model nor a LDD model can reproduce the data. An elliptical model could reproduce the visibility but not the closure phase signal different from 0$^\circ$ or 180$^\circ$, indicating asymmetries.

\begin{figure}[!ht]
	\centering
	\resizebox{\hsize}{!}{\includegraphics{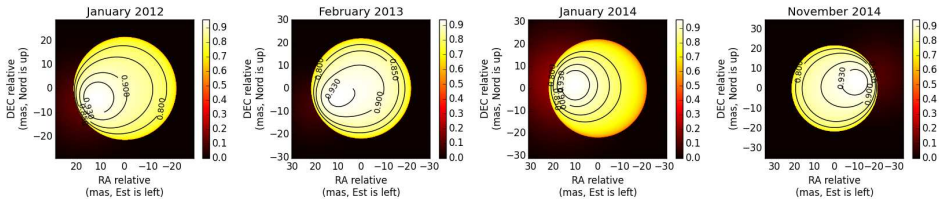}}
	\caption{Intensity maps derived from the best fitted LDD and gaussian hot spot model. North is up and East to the left.}
	\label{Fig:Int_maps}
\end{figure}

We fitted our data using a LDD disk and single gaussian hot spot model, such feature can have a direct effect on the first lobe of the visibility function. We used only visibilities and closure phases below 51~arcsec$^{-1}$. To find the position of the hot spot that gives the absolute minimum of $\chi^2$ we had to use $\chi^2$ maps as the stellar angular size in the model is directly dependent on the characteristics of the gaussian spot. On Fig \ref{Fig:v2_cp_hotspot_2013}, we can see that this model reproduces well the shape of the low spatial frequency data for the 2013 epoch. This is also the case for the other epochs. The intensity images for our four epochs of observations are represented on Fig. \ref{Fig:Int_maps}. The detail of the fitting process is presented in Montarg\`es et al. (submitted to A\&A). This photospheric hotspot is compatible with spectro-polarimetric measurements obtained at the TBL/Narval instrument (Auri\`ere et al. in prep.).

The structure is not reproduced by RHD simulations, indicating probably a missing ingredient. Flux is coming from outside of stellar disk on the intensity maps of the best fitted models. This could be caused by a limb located  convective cell or it could indicate an episodic mass loss event that would be at the origin of a plume emerging from the stellar surface. 

\section{Conclusion}

Near-infrared interferometric observations allowed us to observe the photospheric activity of two nearby RSG. On Antares, the observed signal is matching a RHD simulation up to a resolution of 3~mas. This unprecedented dataset will help bringing numerical constraints on the simulations.

Four years of monitoring of Betelgeuse photosphere revealed a hot spot not reproduced by the simulations and compatible with contemporaneous spectro-polarimetric measurements. Such feature could be at the origin of the episodic mass loss that is creating the inhomogeneous environment of the star (see VLT/NACO and VLT/VISIR observations of \citealt{2009A&A...504..115K,2011A&A...531A.117K}).

\section*{Questions}

\textit{J. Groh:} You mentionned that the hydro simulations do not fit the contrast between the spot and the photosphere, I am wondering what is the physical parameter responsible for setting this contrast.\\
\textit{Answer:} In previous comparison (\citealt{2010A&A...515A..12C}, and \citealt{2014A&A...572A..17M}) we managed to reproduce the contrast as well as the spatial distribution of granulation sizes. However, in the case of PIONIER observations there is clearly something missing in RHD simulations that cannot explain the high contrast between dark and bright structures detected in some particular epochs. This physical ingredients may be the magnetic field or/and the radiative pressure (on molecules or dust). Both are under development.\\

\noindent\textit{A. Moffat :} Are you sure that there are hot spots and not cool spots ? Also: what is the lifetime of the bright hot spots ?\\
\textit{Answer:} To be consistent with the TBL/Narval observations, bright spots are required. However, considering our sampling of the ($u, v$) plane, cool spots and even departure from the disk structure cannot be excluded. The large convective cells seen in infrared evolve on timescales of several years.



\begin{thebibliography}{15}	
	\bibitem[{{Auri{\`e}re} {et~al.}(2010){Auri{\`e}re}, {Donati},
		{Konstantinova-Antova}, {Perrin}, {Petit}, \&
		{Roudier}}]{2010A&A...516L...2A}
	{Auri{\`e}re}, M., {Donati}, J.-F., {Konstantinova-Antova}, R., {et~al.} 2010,
	A\&A, 516, L2
	
	\bibitem[{{Chiavassa} {et~al.}(2011){Chiavassa}, {Freytag}, {Masseron}, \&
		{Plez}}]{2011A&A...535A..22C}
	{Chiavassa}, A., {Freytag}, B., {Masseron}, T., \& {Plez}, B. 2011, A\&A, 535,
	A22
	
	\bibitem[{{Chiavassa} {et~al.}(2010){Chiavassa}, {Haubois}, {Young}, {Plez},
		{Josselin}, {Perrin}, \& {Freytag}}]{2010A&A...515A..12C}
	{Chiavassa}, A., {Haubois}, X., {Young}, J.~S., {et~al.} 2010, A\&A, 515, A12
	
	\bibitem[{{Chiavassa} {et~al.}(2009){Chiavassa}, {Plez}, {Josselin}, \&
		{Freytag}}]{2009A&A...506.1351C}
	{Chiavassa}, A., {Plez}, B., {Josselin}, E., \& {Freytag}, B. 2009, A\&A, 506,
	1351
	
	\bibitem[{{Freytag} {et~al.}(2012){Freytag}, {Steffen}, {Ludwig},
		{Wedemeyer-B{\"o}hm}, {Schaffenberger}, \& {Steiner}}]{2012JCoPh.231..919F}
	{Freytag}, B., {Steffen}, M., {Ludwig}, H.-G., {et~al.} 2012, Journal of
	Computational Physics, 231, 919
	
	\bibitem[{{Haguenauer} {et~al.}(2010){Haguenauer}, {Alonso}, {Bourget},
		{Brillant}, {Gitton}, {Guisard}, {Poupar}, {Schuhler}, {Abuter}, {Andolfato},
		{Blanchard}, {Berger}, {Cortes}, {D{\'e}rie}, {Delplancke}, {di Lieto},
		{Dupuy}, {Gilli}, {Glindemann}, {Guniat}, {Huedepohl}, {Kaufer}, {Le
			Bouquin}, {L{\'e}v{\^e}que}, {M{\'e}nardi}, {M{\'e}rand}, {Morel},
		{Percheron}, {Phan Duc}, {Pino}, {Ramirez}, {Rengaswamy}, {Richichi},
		{Rivinius}, {Sahlmann}, {Schoeller}, {Schmid}, {Stefl}, {Valdes}, {van
			Belle}, {Wehner}, \& {Wittkowski}}]{2010SPIE.7734E...3H}
	{Haguenauer}, P., {Alonso}, J., {Bourget}, P., {et~al.} 2010, in Society of
	Photo-Optical Instrumentation Engineers (SPIE) Conference Series, Vol. 7734
	
	\bibitem[{{Haubois} {et~al.}(2009){Haubois}, {Perrin}, {Lacour}, {Verhoelst},
		{Meimon}, {Mugnier}, {Thi{\'e}baut}, {Berger}, {Ridgway}, {Monnier},
		{Millan-Gabet}, \& {Traub}}]{2009A&A...508..923H}
	{Haubois}, X., {Perrin}, G., {Lacour}, S., {et~al.} 2009, A\&A, 508, 923
	
	\bibitem[{{Hestroffer}(1997)}]{1997A&A...327..199H}
	{Hestroffer}, D. 1997, A\&A, 327, 199
	
	\bibitem[{{Josselin} \& {Plez}(2007)}]{2007A&A...469..671J}
	{Josselin}, E. \& {Plez}, B. 2007, A\&A, 469, 671
	
	\bibitem[{{Kervella} {et~al.}(2011){Kervella}, {Perrin}, {Chiavassa},
		{Ridgway}, {Cami}, {Haubois}, \& {Verhoelst}}]{2011A&A...531A.117K}
	{Kervella}, P., {Perrin}, G., {Chiavassa}, A., {et~al.} 2011, A\&A, 531, A117
	
	\bibitem[{{Kervella} {et~al.}(2009){Kervella}, {Verhoelst}, {Ridgway},
		{Perrin}, {Lacour}, {Cami}, \& {Haubois}}]{2009A&A...504..115K}
	{Kervella}, P., {Verhoelst}, T., {Ridgway}, S.~T., {et~al.} 2009, A\&A, 504,
	115
	
	\bibitem[{{Le Bouquin} {et~al.}(2011){Le Bouquin}, {Berger}, {Lazareff},
		{Zins}, {Haguenauer}, {Jocou}, {Kern}, {Millan-Gabet}, {Traub}, {Absil},
		{Augereau}, {Benisty}, {Blind}, {Bonfils}, {Bourget}, {Delboulbe},
		{Feautrier}, {Germain}, {Gitton}, {Gillier}, {Kiekebusch}, {Kluska},
		{Knudstrup}, {Labeye}, {Lizon}, {Monin}, {Magnard}, {Malbet}, {Maurel},
		{M{\'e}nard}, {Micallef}, {Michaud}, {Montagnier}, {Morel}, {Moulin},
		{Perraut}, {Popovic}, {Rabou}, {Rochat}, {Rojas}, {Roussel}, {Roux},
		{Stadler}, {Stefl}, {Tatulli}, \& {Ventura}}]{2011A&A...535A..67L}
	{Le Bouquin}, J.-B., {Berger}, J.-P., {Lazareff}, B., {et~al.} 2011, A\&A, 535,
	A67
	
	\bibitem[{{Montarg{\`e}s} {et~al.}(2014){Montarg{\`e}s}, {Kervella}, {Perrin},
		{Ohnaka}, {Chiavassa}, {Ridgway}, \& {Lacour}}]{2014A&A...572A..17M}
	{Montarg{\`e}s}, M., {Kervella}, P., {Perrin}, G., {et~al.} 2014, A\&A, 572,
	A17
	
	\bibitem[{{Ohnaka} {et~al.}(2013){Ohnaka}, {Hofmann}, {Schertl}, {Weigelt},
		{Baffa}, {Chelli}, {Petrov}, \& {Robbe-Dubois}}]{2013A&A...555A..24O}
	{Ohnaka}, K., {Hofmann}, K.-H., {Schertl}, D., {et~al.} 2013, A\&A, 555, A24
	
	\bibitem[{{Schwarzschild}(1975)}]{1975ApJ...195..137S}
	{Schwarzschild}, M. 1975, ApJ, 195, 137
	
\end{thebibliography}
\end{document}